\documentclass[aps,prl,twocolumn,groupedaddress,amsmath,superscriptaddress]{revtex4-1}

\usepackage[T1]{fontenc}
\usepackage[latin9]{inputenc}
\usepackage{dcolumn}
\usepackage{amsmath}
\usepackage{amssymb}
\usepackage{amsfonts}
\usepackage{graphicx}
\usepackage{hyperref}
\usepackage{color}
\usepackage{mathrsfs}
\usepackage{multirow}
\usepackage{float}
\usepackage{booktabs}
\usepackage{bm}
\usepackage{dsfont}
\usepackage{txfonts}
\usepackage[english]{babel}
\usepackage{supertabular}

\definecolor{nblue}{rgb}{0.3,0.3,1.0}
\definecolor{ngreen}{rgb}{0.2,0.7,0.2}
\definecolor{nred}{rgb}{0.9,0.1,0}
\definecolor{nblack}{rgb}{0,0,0}

\usepackage[mathlines]{lineno}
\modulolinenumbers[5]

\newcommand{\lt}{\left(}
\newcommand{\rt}{\right)}
\newcommand{\la}{\langle}
\newcommand{\ra}{\rangle}
\newcommand{\lqu}{\left[}
\newcommand{\rqu}{\right]}

\newcommand{\be}{\begin{equation}}
\newcommand{\ee}{\end{equation}}
\newcommand{\ba}{\begin{eqnarray}}
\newcommand{\ea}{\end{eqnarray}}

\newcommand{\bay}{\begin{array}}
\newcommand{\eay}{\end{array}}

\begin{document}

\title{Quantum Enhanced Measurement of an Optical Frequency Comb}
\author{Y. Cai}
\email{caiyin@xjtu.edu.cn}
\address{Laboratoire Kastler Brossel, Sorbonne Universit\'{e}, CNRS, ENS-PSL Research University, Coll\`{e}ge de France, CC74, 4 Place Jussieu, 75252 Paris, France}
\address{Key Laboratory for Physical Electronics and Devices of the Ministry of Education $\&$ Shaanxi Key Lab of Information Photonic Technique, Xi'an Jiaotong University, Xi'an 710049, China}

\author{J. Roslund}
\address{Laboratoire Kastler Brossel, Sorbonne Universit\'{e}, CNRS, ENS-PSL Research University, Coll\`{e}ge de France, CC74, 4 Place Jussieu, 75252 Paris, France}

\author{V. Thiel}
\address{Department of Physics and Oregon Center for Optical, Molecular, and Quantum Science, University of Oregon, Eugene, Oregon 97403, USA}

\author{C. Fabre}
\address{Laboratoire Kastler Brossel, Sorbonne Universit\'{e}, CNRS, ENS-PSL Research University, Coll\`{e}ge de France, CC74, 4 Place Jussieu, 75252 Paris, France}

\author{N. Treps}
\email{nicolas.treps@upmc.fr}
\address{Laboratoire Kastler Brossel, Sorbonne Universit\'{e}, CNRS, ENS-PSL Research University, Coll\`{e}ge de France, CC74, 4 Place Jussieu, 75252 Paris, France}

\begin{abstract}
Measuring the spectral properties of an optical frequency comb is among the most fundamental tasks of precision metrology. In contrast to general single-parameter measurement schemes, we demonstrate here single shot multiparameter estimation at and beyond the standard quantum limit. 
The mean energy and the central frequency of ultrafast pulses are simultaneously determined with a multi-pixel-spectrally-resolved (MPSR) apparatus, without changing the photonics architecture. Moreover, using a quantum frequency comb that consists of multiple squeezed states in a family of Hermite-Gaussian spectral/temporal modes, the signal-to-noise ratios  of the mean energy and the central frequency measurements surpass the shot-noise limit by around $19\%$ and $15 \%$, respectively. Combining our multi-pixel detection scheme and the intrinsic multimode quantum resource could find applications in ultrafast quantum metrology and multimode quantum information processing.

\end{abstract}

\maketitle

\section{Introduction} 
 Optical frequency combs play a fundamental role in many types of precision measurements \cite{Hansch2002, Ludlow2015, Cundiff2003}, including broadband spectroscopy \cite{Thorpe2006, Diddams2007}, absolute frequency determination \cite{Diddams2001,Rosenband2008}, optical clocks \cite{Ye2011, Katori2011}, and time-distance synchronization \cite{Coddington2009, VanDenBerg2012}.  Measuring the spectral properties of an optical pulse is thus an important part of precision metrology \cite{Bartels2004, Haus1990, Haus1993}. The quantum-limited sensitivity for such measurements is dictated by the noise fluctuations present in a well-defined spectral mode  \cite{Haus1990, Schmeissner:2014uv}, and the time and spectral separation has been performed experimentally \cite{Silberhorn2018}. For instance, the central frequency of optical pulses corresponds to the derivative of the line shape to be measured \cite{Pu2012, Pinel2012,  thiel2017}. The measurement precision and signal-to-noise ratio in these applications are generally limited by photon number fluctuations, which scale as $\sqrt{N}$, where $N$ is the number of photons in the beam to be detected \cite{Bachor2004}. Optimally engineered squeezed states of light may be utilized to achieve a sensitivity beyond this quantum shot-noise limit \cite{Caves1981}, which has been widely applied in various measurements, such as laser interferometers \cite{Xiao1987, Grangier1987, Eberle2010, Steinlechner2018}, gravitational wave interferometry \cite{LIGO2011, LIGO2013, Grote2013, Tse2019},  optical magnetometry \cite{Wolfgramm2010,Li2018}, laser beam pointing \cite{Pooser2015, Treps2003}, biological sensing \cite{Taylor2013}, distributed phase sensing\cite{guo2019}, \emph{etc}.

However, in order to characterize a physical system with several physical quantities, one often needs to measure  multiple related parameters many times while modifying the corresponding set-up, which lacks flexibility. Here, we introduce a multimode approach for parallel estimation of multiple orthogonal parameters of a light field, as well as quantum-enhanced metrology with incorporating an intrinsic multimode squeezed light. In this letter, we demonstrate a shot-noise limited measurement of multiple parameters characterizing the field of optical pulses of a frequency comb. To realize such parallel multiple parameter estimation, a multi-pixel-spectrally-resolved (MPSR) detector is developed. Post-processing the multichannel data of a single shot measurement, the central frequency and the mean energy of light field are revealed. Also using this spectrally-resolved apparatus, the full covariance matrix of a quantum frequency comb is reconstructed by simultaneously measuring all the spatially separated spectral components.
Furthermore, the signal-to-noise-ratio (SNR) of the central frequency and the mean energy measurements are enhanced with the appropriate use of an ultrafast squeezed light pulse \cite{Roslund:cb, Cai2017}. We thus demonstrate multiple parameter estimation of an optical frequency comb, \emph{i.e.} beyond the shot noise limit, without changing the photonics architecture.

\section{The Quantum Cram\'{e}r-Rao Bound of Spectral Measurements} 
 Let us consider the complex electric field of a single pulse of light, $\mathcal{E}\lt t\rt$, as the product of a mean amplitude and a specific pulse shape, $\mathcal{E}\lt t\rt=A_t u\lt t\rt\textrm{exp}\lt i\omega_0t\rt$, where $u\lt t\rt$ represents a specific time mode (\emph{i.e.} a single pulse shape),  $\omega_0$ is the central frequency, and $A_t$ is the complex amplitude. This pulse can be a single one or part of a train of pulses, such as an optical frequency comb, and this does not influence the following derivation as soon as the measurement device has a spectral resolution much lower than the comb repetition rate. The optical field can then simply be expressed in the frequency domain as
\be
\mathcal{E}(\omega)=\mathcal{E}_0 E\lt\omega\rt=\mathcal{E}_0 \sqrt{N}u\lt\omega\rt,
\label{spectrum}
\ee
where $\mathcal{E}_0$ is a normalisation constant chosen so that $N$ represents the mean photon number, and $u(\omega)$ is the normalized spectral amplitude. For a mean field mode with a gaussian spectral shape, we have:
\be
u\lt\omega\rt=\sqrt{\frac{1}{\sqrt{2\pi}\Delta \omega}}\textrm{exp}\lqu\frac{-\lt\omega-\omega_0\rt^2}{4\Delta\omega^2}\rqu,
\ee
where $\omega_0$ is the central frequency, $\Delta \omega$ is the spectral width and $\int_{-\infty}^{\infty} u\lt\omega\rt u^*\lt\omega\rt\textrm{d}\omega$ is normalized to be unity. Thus to define such an optical pulse,  three parameters, $N$, $\omega_0$ and $\Delta\omega$, need to be characterized. 

\begin{figure}[ht]
\centering
\includegraphics[width=85mm]{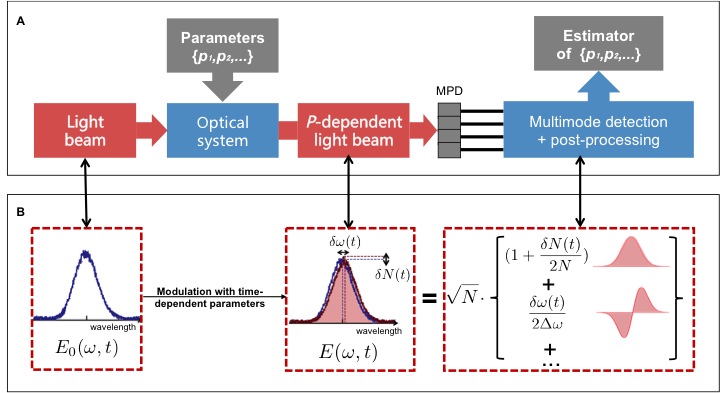}
\vspace{5mm}
 \caption{A. Parameter estimation with a multimode detection scheme. A light beam carrying the parameters is interrogated with a multi-pixel detector (MPD), and via post-processing the associated optical modes, multiple orthogonal parameters can be simultaneously estimated. B. Modal decomposition of a noisy optical pulse: physical noise parameters can be associated with different time/frequency modes. We represent here photon number and central frequency noise modes. Measurement of the associated modes can lead to Cram\'er-Rao bound limited sensitivity}.
 \label{modes}
 \end{figure}

Let us now consider that a small variation of the central frequency, $\delta \omega$, and of the mean energy,  $\delta{\mathcal{N}}=\mathcal{E}_0^2\delta N$ are present within the optical pulse. Then the electric field can be expressed as:
\be
E_{\delta\omega,\delta N}\lt\omega\rt
\approx\sqrt{N}u\lt\omega\rt+\frac{1}{2\sqrt{N}}\delta{N}u\lt\omega\rt+\frac{\sqrt{N}}{2\Delta\omega}\delta\omega u_\textrm{d}\lt\omega\rt.
\label{mode}
\ee
 
The parameters corresponding to a displacement $\delta \omega$ of the central frequency and $\delta \mathcal{N}$ of the mean energy are carried by a specific mode or pulse shape. These normalized modes are respectively given by $u_\textrm{d}\lt\omega\rt=2\Delta\omega\partial u\lt{\omega}\rt/\partial\omega$ and  $u\lt\omega\rt$. This approach is very general to any parameter, and these mode-dependent parameters could be  estimated with a multimode detection scheme, as illustrated in Fig. \ref{modes}. An optical pulse perturbed by any type of noises can always be decomposed on a basis of unperturbed orthogonal modes, which carry corresponding time-dependent coefficients. These coefficients are associated with the variation of specific physical parameters. Their noise properties is ultimately governed by the quantum vacuum fluctuations in the corresponding modes, leading to the shot noise limit in the measurement of these parameters. Hence, using squeezed light in some of these modes allows for a measurement with a precision that surpasses the shot noise limit~\cite{Treps2003, Treps2005, Lamine2008}.

To describe the quantum limit in the measurement of these quantities, a full quantum representation of light is taken. Hence,  the electric field operator is expressed as $\hat{E}\lt \omega\rt =\sum_i \hat{a}_i u_i\lt\omega\rt$, where $\hat{a}_i$ is the annihilation operator in the spectral mode $u_i\lt\omega\rt$, and where we did omit the constant $\mathcal{E}_0$ for simplicity.  In the case of coherent state illumination, the Cram\'{e}r-Rao bound for the parameters $\delta N$ and $\delta \omega$ of \eqref{mode} is given by~\cite{Pinel2012, Pu2012, thiel2017}
\ba
(\delta N)_\textrm{SQL}&=&\sqrt{N},\\
(\delta \omega)_\textrm{SQL}&=&\frac{\Delta \omega}{\sqrt{N}}.
\label{band}
\ea

In the more general case where the noise of the mode carrying the parameter is not at the shot noise level, but still Gaussian, the Cram\'{e}r Rao bound is constrained by that noise:
\ba
\delta N&=&\sqrt{N} \sqrt{\langle \Delta^2 \hat{x} \rangle}\label{eq:dNsens},\\
\delta \omega&=&\frac{\Delta \omega}{\sqrt{N}}\sqrt{\langle \Delta^2 \hat{x}_d \rangle},
\label{eq:dwsens}
\ea
where $\hat{x}$ and $\hat{x}_d$ are the optical amplitude quadrature of modes $u$ and $u_d$ respectively, defined as $\hat{x}=\hat{a}+\hat{a}^\dag$ and $\hat{x}_d=\hat{a}_d+\hat{a}^\dag_d$, and $\Delta^2 \hat{x}$ and $\Delta^2 \hat{x}_d$ are the noises in the associated modes.  These limits are the ones we are aiming at using our experimental scheme.

\section{Model of the MPSR Detection System}

\begin{figure}[htb]
\centering
\includegraphics[width=85mm]{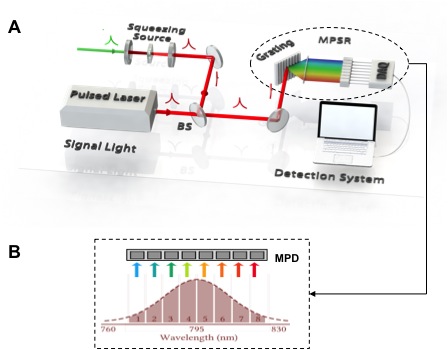}
  \vspace{5mm}
  \caption{A. Experimental setup for the simultaneous multimode measurement of frequency and energy fluctuations within a laser field. Both parameters are modulated within the pulsed laser cavity. A multimode squeezing resource is then employed which allows enhancing the measurement of either quantity beyond the standard quantum limit. This is accomplished by mixing the squeezer with the laser field via a strongly reflective beamsplitter. The resultant synthetic beam is spectrally dispersed and imaged onto a photodiode array after passing a micro-lenses array. The photocurrents corresponding to each of the eight pixels are transferred to a computer for post-processing. BS: Beam splitter, 10/90; Grating has 93$\%$ efficiency; MPSR: homemade multi-pixel-spectrally-resolved detector. B.  The corresponding spectral bins of of equivalent width are simultaneously interrogated with a eight-pixel photodiode array. MPD: Multi-pixel detector. }
  \label{setup}
  \end{figure}

We now describe the model of the multi-pixel-spectrally-resolved (MPSR) detection which is employed to reach the Cram\'er-Rao bound, as introduced in the previous section. This detection system is represented in Fig. \ref{setup}, and has for input, in our case, a train of light pulses which carry the information to be extracted. These pulses are diffracted on a grating and the different color components are then detected simultaneously by a one-dimension photodiode array.  Hence, the modes effectively detected by the multi-pixel array, that we call the pixel modes and write also $\{u_i\}$ for simplicity, are the normalized spectral slices of the mean field of the intense beam shining on the detector pixels. Thus, these pixel modes, $\{u_i\}$, define the measurement basis. 

In order to recover the mode which carries a given parameter, for instance the central frequency, one can implement a real linear basis change on these measured photocurrents. Let us call $\{v_s\}$ the mode we want to recover, associated with annihilation operator $\hat{a}_s$. We approximate this mode from the pixel modes:
\be
\hat{a}_s\simeq \hat{a}_m = \frac{\sum_i m_i\hat{a}_i}{\eta},
\label{signal mode}
\ee
where the projection coefficients, $m_i=\int u_i^*(\omega)v_s(\omega)\textrm{d}\omega$  and the detection efficiency,  $\eta=\sqrt{\sum_i m_i^2}$. We assume, as is the case both in the theoretical description and in the experiment, that all the $m_i$ are real, meaning that the spectral phase of the mode $v_s$ is the same as the one of the pixel modes. Note that one can easily include this possible phase variation in the calculation, but we omit it for simplicity .
Hence, if the detection efficiency $\eta$ is equal to 1, the mode $\hat{a}_s$ is perfectly recovered and the measurement sensitivity exactly reaches the Cram\'{e}r Rao bound~\cite{Treps2005}. In the general case, the quality of the approximation between the signal mode $\hat{a}_s$ and the measured mode $\hat{a}_m$ depends ultimately on the number of pixels and on the filling factor of these pixels onto the photodiode array relative to the pixel modes of the optical spectrum to be measured.

More specifically, because we aim at measuring a variation of the mean field as defined in \eqref{mode}, we can consider the small variations of the measured intensity on each pixel mode relative to the mean power. As the mean field is intense, one can write that the variation of $\hat I_i = \hat a^\dagger_i\hat a_i$ is equal to $\delta \hat I_i = \alpha_i\delta\hat x_i$ where $\alpha_i$ is the amplitude of the field in the $i$th pixel mode $\hat a_i$, considered as real without loss of generality, and $\delta \hat x_i = \hat x_i - \langle \hat x_i \rangle$ is the fluctuation quadrature operator. Hence, post-processing the measured intensity using \eqref{signal mode} one can directly reconstruct $\delta \hat x^m = \frac{1}{\eta}\sum_i m_i \frac{\delta I_i}{\alpha_i}$ and access the fluctuations of mode $\hat a_m$. In the case where $\eta=1$, the multipixel detection system is able to estimate the parameter carried on a specific mode, reaching the Cram\'{e}r Rao bound of \eqref{eq:dNsens} and \eqref{eq:dwsens}.
If the modes  are squeezed on the amplitude quadrature ($(\delta\hat{x}^m)^2<1$), the sensitivity increases beyond the shot noise limit, allowing for a smaller value of the corresponding parameters to be measured. The case $\eta<1$ is equivalent to a reduced efficiency of the detection, and lead to a corresponding reduced sensitivity.

Furthermore, using conventional single-pixel detection methods, such orthogonal parameters can't be simultaneously measured. However, using the MPSR, one may choose several on-demand spectral modes to characterize and apply the corresponding post-processing simultaneously. The MPSR allows arbitrary real basis change on the original multi-pixel data. Thus, this methodology enables parallel parameter estimation via post-processing without changing the optical set-up.  

\section{EXPERIMENTAL SET-UP and shot-noise limited measurement }
 \begin{figure}[htb]
\centering
\includegraphics[width=70
mm]{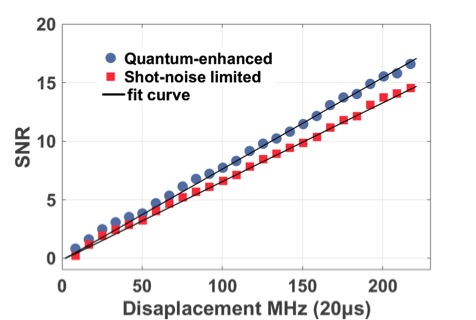}
  \vspace{5mm}
  \caption{Measurement of the central frequency displacement that is induced inside the laser cavity. The blue and the red lines are the SNRs for the shot-noise limited and quantum-enhanced cases, respectively. All the measurements are implemented at 1.5 MHz modulation sideband where all the classical noises could be ignored. All the data are only corrected for electric dark noise. The measurement time is taken to be 20 $\mu$s for each measurement, which is defined by the bandwidth (50 kHz) of the low pass filter. The error bar for each point above is less than 2$\%$,  as the system is well locked and many events are used in the measurements. These quantities of the error bars are small, and are thus not drawn for clarity. 
}
\label{SNR}
  \end{figure}

Our experimental set-up is displayed in Fig. \ref{setup}. 
A train of $\sim100~\textrm{fs}$ optical pulses centered at 795 nm with a repetition rate of 76 MHz propagates through a weakly transmitting beam splitter.  
In practice, the center wavelength of the spectrum is shifted by modulating at a high frequency the tilt of a mirror inside the pulsed laser cavity. The modulation is done at $f_m=1.5$ MHz in order to avoid the technical noise of the laser field and ensuring that the measurement is done at the shot noise limit. The optical beam is spectrally dispersed by a high efficiency grating and imaged onto a eight-pixel photodiode array. To avoid the gaps in between pixels of the photodiode array, we use a micro-lenses array to focus the eight color components on each pixel. The low-noise amplifiers of the photodiodes have a bandwidth of about 10 MHz. The photocurrents from the individual pixels are then individually demodulated at the modulation frequency $f_m$ and fed into a data acquisition system after passing through a 50 kHz low-pass filter.

Once all the photocurrents are acquired by the computer, we can calculate the signal and noise of different spectral modes as defined in \eqref{signal mode}. The coefficients $m_i$ in \eqref{signal mode} are calculated from the low-frequency output of the MPSR, which gives the spectral amplitude of the mean field, and the predicted modes associated with each parameter. In order to retrieve the sensitivity, we vary the modulation depth in the laser cavity and calculate for each modulation the signal to noise ratio using 1000 data points at a sampling rate of 20 kHz. It is represented in fig. \ref{SNR} with the particular example of the derivative mode associated with the central frequency fluctuations. The blue curve corresponds to the obtained sensitivity for coherent input light.

In order to calibrate the sensitivity and the measurement, we evaluate its properties. On the photodiode array, the field is spread over a detected bandwidth of $8.8$ nm full width half maximum and contains a photon flux of $N_p\simeq 4\times10^{16}$ photons/second (which corresponds to $\sim$10 mW of total optical power). These experimental values lead to a shot noise limited sensitivity as defined in \eqref{eq:dwsens} of $\delta\omega_\textrm{SQL} \approx $ 55.7 kHz/$\sqrt{\textrm{Hz}}$, which is much smaller than the 76 MHz spacing between the comb tooth. Considering the quantum efficiency of the photodiodes, the optical losses and limited number of pixels that induce a measured non-perfect mode mismatch with the ideal spectral mode, the global efficiency in intensity $\eta^2$ is evaluated as 70$\%$, and the practical sensitivity is thus $\delta\omega_\textrm{meas.} \approx $ 66.5 kHz/$\sqrt{\textrm{Hz}}$. The measurement time, induced by the low-pass filter, is 20 $\mu$s, thus experimentally,  the actual  sensitivity presented in Fig. \ref{SNR} is calculated to be 14.9 MHz for each measurement event.

\textit{The quantum frequency comb and Quantum-enhanced spectrometer} In this section, we show how the MPSR detector can be utilized with multimode squeezed vacuum to simultaneously measure orthogonal parameters of the optical field with a precision that surpasses the shot noise limit.

 \begin{figure}[h]
\centering
\includegraphics[width=80mm]{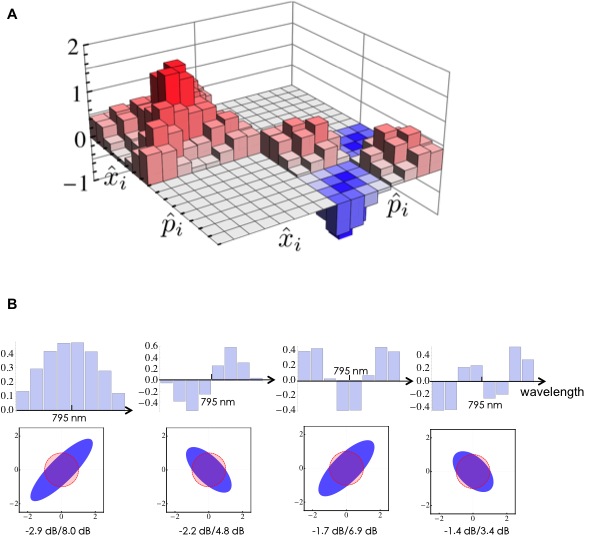}
 \vspace{5mm}
\caption{
	A. 8-partite full covariance matrix of the multimode quantum resource via multi-pixel homodyne measurement. All the amplitude or phase quadrature of different color components, $\hat{x_i}$ or $ \hat{p_i}$, are simultaneously interrogated, and the spectrum of the local oscillator is divided into 8 frequency bands of equivalent width. The covariance matrix elements are defined as $\frac12\left\langle {{{\hat\zeta }_i}{{\hat \zeta }_j} + {{\hat \zeta }_j}{{\hat \zeta }_i}} \right\rangle - \left\langle {{{\hat \zeta }_i}} \right\rangle \left\langle {{{\hat \zeta }_j}} \right\rangle$, where $\hat{\zeta}_i=\hat{x}_i, \hat{p}_i$. The shot noise contribution has been subtracted from the diagonal for better visibility, and noise level is normalized to vacuum noise (\emph{i.e.} shot noise is equal to 1). Note that our squeezer presents no amplitude-phase cross correlation, hence the corresponding parts in the full covariance matrix are all zero. 
	B. The squeezing values of the four leading spectral modes (up) and the corresponding squeezing ellipse in quadrature phase space. The red circles represent the shot noise, and the odd-order and even-order modes are squeezed in amplitude and phase quadrature, respectively. Both (A) and (B) were only corrected for electrical dark noise. 
	}
\label{8covariance}
\end{figure}

Multimode quantum light is generated by an optical parametric oscillator synchronously pumped by the second harmonic of a femtosecond mode-locked laser, as shown in Fig. \ref{setup}. The quantum state consists of multimode squeezed vacuum on a basis of time/frequency modes that resemble closely the modes obtained by expansion of the field in \eqref{mode}\cite{Roslund:cb}. To characterize that quantum state, we use two MPSR detectors in a spectrally-resolved homodyne detection scheme to simultaneously record the quadrature operators in the frequency band basis. We use this information to reconstruct the eight-partite covariance matrix of the quantum frequency comb, both in amplitude and phase, as depicted in Fig. \ref{8covariance}A. 
Note that the total interrogation time to obtain the full covariance matrix is a few seconds with vacuum squeezing locked on the amplitude or phase quadrature (see appendix for details). This measurement technique is also promising for quantum information processing, which requires simultaneous interrogation of the various entangled parties \cite{Su2013, Furusawa2011, Menicucci2008, Andersen2015}.

The full covariance matrix contains all the quantum correlations of the multimode gaussian quantum resource. Using Bloch Messiah decomposition \cite{Braunstein2005} we extract from the covariance matrix a set of orthogonal squeezed modes, which happen to be very similar to a set of Hermite-Gaussian spectral shapes in the frequency domain (corresponding to pulse shapes in the time domain). The leading four squeezed eigenmodes are presented in Fig. \ref{8covariance}B., which are respectively squeezed by -2.9 dB, -2.2 dB, -1.7 dB and -1.4 dB, corrected for electrical dark noise.  Note that the odd order eigenmodes are squeezed on the amplitude while the even order ones are squeezed on the phase quadrature. Here the squeezing levels are mainly limited by the quantum efficiency of the photodiode array($\sim~80\%$ quantum efficiency, Hamamatsu S8558), but ones could see that the mean field mode and the derivative mode are close to the leading two eigenmodes which are both significantly squeezed.

In order to implement  the quantum-enhanced measurement, the quantum frequency comb generated by the OPO is then combined with the optical field coherently on a weakly transmitting beamsplitter. In this way, the generated beam carries the mean field of the original comb, yet with the quantum fluctuation of the quantum frequency comb. Therefore,
the synthetic beam consists on a multimode optical field that carries the parameters to be measured with a noise that is either below or above shot noise because of quadrature squeezing. 
In practice, as the beamsplitter reflectivity is about 90\%, the squeezed state undergoes about 10\% losses.  
Experimentally multiple optical phases have to be locked together. The OPO has to be seeded in order to be locked on amplification, \emph{i.e.}, the resulting eigenmodes are squeezed on the amplitude quadrature. Additionally, the strong signal field and the squeezed vacuum have to be locked together, this is achieved by locking the signal to the seed. The measurement with quantum vacuum is then performed employing a fast mechanical shutter which blocks the seed beam whilst holding the electrical locks, allowing for a 100 ms measurement window before the shutter opens and the locks are resumed. Once data is acquired, the analysis is then performed in the exact same way as in the previous section.

Reconstructing the mode associated with central frequency fluctuations $u_d$, we see that the SNR is enhanced by $\sim 15\%$ with respect to the standard shot-noise sensitivity. As seen in Fig. \ref{SNR}, we recover a higher response to the modulation depth compared to the previous case. By computing the ratio between the slopes of the blue (shot noise limited) and red (quantum enhanced) curves, the quantum-enhanced sensitivity is  $\delta\omega_{Quan}\approx$ 57.8 kHz/$\sqrt{\textrm{Hz}}$, thus with 20 $\mu$s measurement time, the actual quantum-enhanced sensitivity is 12.9 MHz.

\begin{figure}[ht]
\centering
\includegraphics[width=85mm]{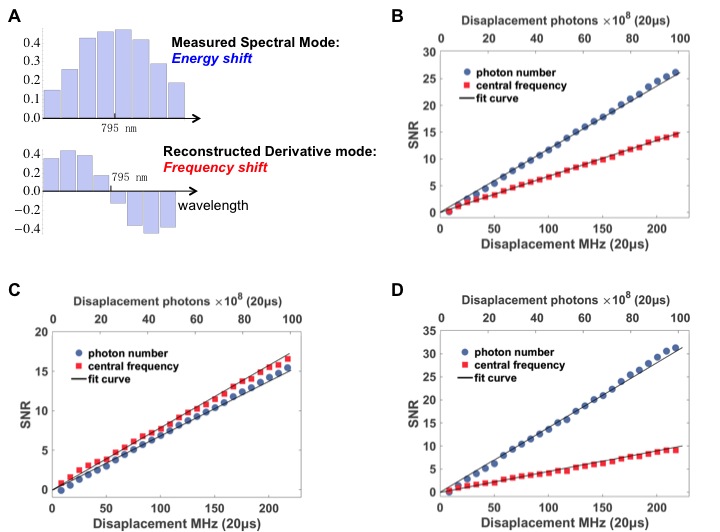}
\caption{
	A. Experimental detection modes corresponding to the mean-field (top) and the derivative mode (bottom), which is reconstructed and orthogonalized to the mean-field.
	B-D. SNRs for both mean-field  (blue curves) and derivative mode (red curves) whilst increasing the center frequency modulation depth, B. when the other port of the beamsplitter is vacuum, C. quantum frequency comb with a squeezed derivative mode, and D. quantum frequency comb with a squeezed mean-field mode. As the system is well locked and many events are used in the measurements, these quantities of the error bars for each point of SNRs above are small (less than 2$\%$), and are thus not drawn for clarity.   
}
\label{multimodefre}
\end{figure}

The MPSR is a flexible platform that allows for simultaneous extraction of multiple orthogonal modes by post-processing. In addition to quantifying the frequency fluctuations, variations in the mean energy of the light field are simultaneously obtained by reconstructing an alternative photocurrent superposition (i.e., the sum of the individual photocurrents which corresponds to the mean-field mode). In practice, the modulation of the center frequency within the laser cavity also causes an energy variation. According to \eqref{mode}, both variations in energy and center frequency can be retrieved by post-processing the mean-field mode and the corresponding derivative mode, respectively.  As seen in Fig. \ref{multimodefre}, while the sensitivity for determining frequency deviations is enhanced with a squeezing resource, the sensitivity for the measurement of energy fluctuations is degraded with respect to the shot noise limit. This is due to the fact that the modes corresponding to these two spectral structures are squeezed in opposite quadratures (i.e., the derivative mode is squeezed while the mean-field mode is anti-squeezed).  However, acting on the locking system one can change which parameter is to be squeezed, and switch to a synthetic beam whose energy parameter estimation is enhanced. The corresponding shot noise limited sensitivity of mean energy shifts is $\delta N_{SQL} \approx$  $2 \times 10^8$ photons/$\sqrt{\textrm{Hz}}$, and with considering the global efficiency, the practical sensitivity is $\delta N_{mesa} \approx$  $1.7 \times 10^8$ photons/$\sqrt{\textrm{Hz}}$. According to the SNR ratio in Fig. \ref{multimodefre}, the quantum enhanced sensitivity for the mean energy shift is $\delta N_{Quan} \approx$$1.4 \times 10^8$ photons/$\sqrt{\textrm{Hz}}$, which is $\sim 19\%$ improved compared to the one with shot noise limited. For a measurement time of 20 $\mu$s, the actual sensitivities are $3.8 \times 10^8$ photons and $3.1 \times 10^8$ photons for shot-noise limited and quantum-enhanced, respectively.

\textit{Conclusions and Discussions}
The present work provides a proof of principle for the ability to exceed the standard quantum limit in the measurement of frequency and energy fluctuations within an optical frequency comb.  Importantly, the use of a multiplexed detection device is general and enables the simultaneous estimation of multiple parameters characterizing the light pulses. Both displacement of central frequency and mean energy of pulses are measured in parallel with the shot-noise limited; furthermore, jointly applying the quantum frequency comb, which is intrinsic multimode squeezed, we achieve the sensitivity of both parameters, i.e, the displacement of the mean energy and the central frequency, beyond the shot-noise limit by 19$\%$ and 15$\%$, respectively.

Notably, appropriate linear combinations of the multiple photocurrents of the wavelength-multiplexed detection have also revealed variations in the field bandwidth, temporal jitter, and overall phase \cite{Schmeissner:2014uv}. This fact suggests a tremendous flexibility in the use of multimode detection for the interrogation of multidimensional light fields.  Moreover, multimode measurement based quantum computing and multipartite quantum secure communications could also be implemented with a multi-pixel-spectrally-resolved homodyne detection scheme \cite{Ferrini2013, CaiArxiv2019}. Applying the present multi-pixel detection scheme together  with a controllable non-Gaussian operation such as photon subtraction \cite{RaNP} is a promising way to realize quantum advantage.

\textbf{Funding Information} The French National Research Agency projects COMB and SPOCQ, National Key R$\&$D Program of China (Grants No. 2017YFA0303700), the National Natural Science Foundation of China (Grants No. 61975159 and No. 11904279),  and the National Science Foundation of Jiangsu Province (Grant No. BK20180322).

\textbf{Acknowledgments} C. F. and N. T. acknowledge financial support of the Institut Universitaire de France.
Y.C. thanks the National Key R$\&$D Program of China, the National Natural Science Foundation of China, and the National Science Foundation of Jiangsu Province, as well as the basic research project and the young researcher program in Xi'an Jiaotong University.

\appendix*
\setcounter{equation}{0}
\renewcommand\thefigure{A\arabic{figure}}
\renewcommand\theequation{A\arabic{equation}}
\renewcommand\thetable{A\arabic{table}}
\subsection*{Appendix: Multi-pixel homodyne detection}

\begin{figure*}[!htb]
\centering
\includegraphics[width=150mm]{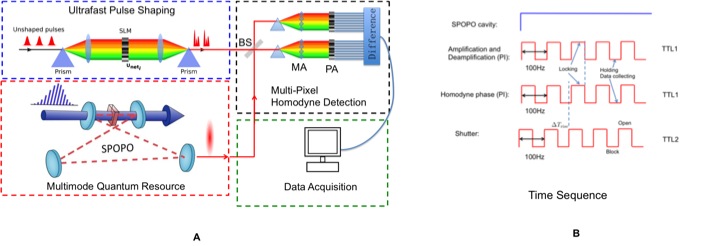}
 \vspace{5mm}
\caption{(A). Simultaneous measurement regime with multi-pixel Homodyne detection. The multimode quantum resource is generated via a synchronously pumping an optical parametric oscillator (SPOPO). The local beam of homodyne detection could apply pulse shaping technique to access on-demand pulse shapes, which is not yet used for constructing covariance matrix. The two beams, going through a balanced  beam splitter (BS), are dispersed by a pair of dispersive optics, and each color injects on the corresponding pixel of the the photodiode array (PA). To eliminate the loss induced by the gaps between the pixels of PA, a pair of  micro-lens arrays (MA) is set in front of the PA.    (B). Time sequence for locking and data acquisition. }
\label{MuliHomo2}
\end{figure*}

The multipartite frequency structure of ultrafast pulse trains provides a rich platform for the generation of squeezed modes of various spectral shapes \cite{Roslund:cbap}. The second harmonic of a 76 MHz pulse train delivering $\sim120$ fs pulses centered at 795 nm synchronously pumps a below-threshold optical parametric oscillator (OPO). The squeezed vacuum output of this Synchronously Pumped OPO (SPOPO) has been demonstrated to consist of a multimode squeezed state in which the spectral structure of each mode closely approximates a Hermite-Gaussian progression  \cite{Roslund:cbap}. For instance, the second squeezed mode of the series provides a close approximation to the derivative of the original spectral field, which is the optimal mode for identifying frequency variations \cite{Pinel2012ap, Schmeissner:2014uvap}. 

The multimode quantum correlations of the quantum resource are measured via multi-pixel balanced homodyne detection. This spectrally-resolved apparatus provides many individual homodyne detecting different colors of pulses in the same time, which enables to collect the quantum correlations among all channels simultaneously. The full covariance matrix of the multimode gaussian state can then be reconstructed  with the measured quantum correlations. Importantly, the simultaneous homodyne measurement is required in multimode quantum information processing \cite{Ferrini2013ap} and allows for parallel parameter estimation \cite{Pinel2012ap}.  

As seen in Fig. \ref{MuliHomo2}(A), the multimode quantum resource  \cite{Roslund:cbap, Cai2017ap}, and the local oscillator are balanced mixed with the beam splitter. 
Different from general  homodyne detection, after the $50/50$ beam splitter, the two balanced arms pass through an optical dispersion device, and all the frequency components  are thus simultaneously interrogated via the homemade multi-pixel detection apparatus. To reconstruct the full covariance matrix, amplitude and phase quadrature correlations are collected by two single measurements, while locking the relative phase between the signal and the local light at 0 and $\pi/2$, respectively.

Experimentally, for dispersing the light, a pair of optical gratings with the efficiency of 93$\%$ are used . The homemade multi-pixel apparatus has $\sim 80\%$ detection efficiency, 10 MHz detection bandwidth, and commercial photodiode arrays ($\sim80\%$ quantum efficiency, Hamamasu S8558) are used, as well as microlensarray is applied to focus the dispersed light onto the sensing pixels of the photodiode arrays. The homodyne visibility is $94\%$. 
The cumulative loss of the system is  $25\%$, which limits the measured squeezing level. The data is interrogated with a vacuum squeezing locking. As seen in  Fig. \ref{MuliHomo2}(B), the time sequence is trigged with a two-channel  high precision signal generator, which generates TTL1 and TTL2 with a fixed relative time difference of a few milliseconds. A fast mechanical  shutter is used to control the seed beam of SPOPO, and its on-off frequency is set to be 10--100 Hz. The both locking of the cavity and amplification (deamplification) is on with seed beam, and the covariance data of SPOPO is collected while the seed beam is blocked and the locking is hold.

To reconstruct the covariance matrix, the quantum correlations are measured via the multipixel homodyne detection. All the difference signal of each pair of the frequency pixels in the multi-pixel homodyne detection are measured simultaneously, and the difference of the $i$th pair of detector pixels represent the quadrature value in the $i$th frequency band,  expressed as below,
\be
\hat{I}^-_i=2\vert\alpha_i\vert^2 \hat{o}_i,
\ee
where $\alpha_i$ is the amplitude of the $i$th frequency band of the local oscillator, and $\hat{o}_i$ represents the corresponding quadrature of the signal field. The covariance matrix with amplitude and phase quadrature correlations are reconstructed by calculating the variances of combined difference signals.
To construct the full covariance matrix, the correlations terms are defined,
\be
\langle \hat{o}_i \hat{o}_j\rangle=\frac{\la\hat{I}^-_i\hat{I}^-_j\ra}{4\vert\alpha_i\vert^2\vert\alpha_j\vert^2}.
\ee

Importantly, the first and second squeezed modes of the series provides a close approximation to the mean field mode and the derivative mode, which identify energy and central frequency variations \cite{Roslund:cbap, Pinel2012ap}. Hence,  combining the multimode squeezed states and the multichannel detector allows for parallel estimation of corresponding parameters beyond the standard quantum limit.



\end{document}